\documentclass[numbered]{trbunofficial}
\usepackage{booktabs}
\usepackage{tabularx}
\usepackage{array}
\usepackage{multirow}
\usepackage{graphicx}
\usepackage{listings}
\usepackage{xcolor}
\usepackage{float}
\usepackage[section]{placeins}

\definecolor{lstkw}{HTML}{1F4E79}
\definecolor{lstcmt}{HTML}{7B8794}
\newcommand{\AwCommit}{\texttt{e53832e8}}

\newcommand{\NTU}{770}
\newcommand{\NUnits}{4{,}673}
\newcommand{\NUnitsSC}{1{,}106}
\newcommand{\NUnitsStrat}{120}
\newcommand{\NUnitsEval}{161}
\newcommand{\NUnitsKnown}{41}
\newcommand{\NParseFail}{1}
\newcommand{\PctSC}{23.7}

\newcommand{\NClasses}{18}
\newcommand{\NClassAV}{7}
\newcommand{\NClassAmp}{11}

\newcommand{\NEvidenceLinks}{416}
\newcommand{\NDistinctCWE}{18}

\newcommand{\NCVERetrieved}{393}
\newcommand{\NCVERetained}{372}
\newcommand{\NCorpusIndexed}{374}
\newcommand{\NCorpusNVD}{372}
\newcommand{\NCorpusAdv}{2}
\newcommand{\EmbedModel}{\texttt{nomic-embed-text}}
\newcommand{\EmbedDim}{768}
\newcommand{\RagTopK}{6}
\newcommand{\RagMinSim}{0.35}

\newcommand{\NKnownIssues}{61}
\newcommand{\NKnownHigh}{17}
\newcommand{\NKnownMed}{44}
\newcommand{\NCommitsScanned}{7{,}561}
\newcommand{\NBulkRejected}{3}
\newcommand{\NBulkRejectedLocs}{159}

\newcommand{\CodeQLPack}{\texttt{codeql/cpp-queries@1.5.10}}

\newcommand{\AFLVer}{4.00c}
\newcommand{\NCustomQL}{4}
\newcommand{\NCustomSemgrep}{8}
\newcommand{\DBCovUnits}{161}

\newcommand{\DBCovFiles}{97}

\newcommand{\NCppcheckFindings}{1{,}301}

\newcommand{\NRunsTotal}{1{,}318}

\newcommand{\PCodTaxMapped}{82.3}

\newcommand{\PCodZsMapped}{0.0}

\newcommand{\POssTaxMapped}{95.9}

\newcommand{\POssZsMapped}{2.2}

\newcommand{\NOssMedFullUnits}{15}

\newcommand{\FuzzBudget}{900}
\newcommand{\NRepairRounds}{2}
\newcommand{\NConfProcessed}{187}
\newcommand{\NConfirmed}{0}
\newcommand{\NInconclusive}{187}
\newcommand{\NHarnessBuilt}{1}
\newcommand{\NHarnessBuildFail}{173}

\newcommand{\NNoCrash}{13}
\newcommand{\NLLMSubmitted}{187}
\newcommand{\NBaselineSubmitted}{0}

\newcommand{\MatchTol}{10}
\newcommand{\NGroundTruth}{46}

\newcommand{\AblZsR}{0.7391}

\newcommand{\AblTaxR}{0.8478}

\newcommand{\AblRagF}{0.2365}

\newcommand{\AblFullF}{0.2783}

\newcommand{\NullIters}{2{,}000}
\newcommand{\NullPFloor}{0.0005}
\newcommand{\NullCodZsObs}{0.696}
\newcommand{\NullCodZsExp}{0.327}

\newcommand{\NullCodTaxObs}{0.761}

\newcommand{\NullOssZsObs}{0.217}

\newcommand{\NullOssTaxObs}{0.435}

\newcommand{\PctCodTaxRecall}{76}

\newcommand{\NLLMConditions}{8}

\newcommand{\NToolProducers}{5}
\newcommand{\MinLLMLift}{+0.142}
\newcommand{\MaxLLMLift}{+0.379}

\newcommand{\NBaseAll}{2{,}300}
\newcommand{\NBaseInUnits}{77}
\newcommand{\NBaseUnitsTouched}{36}

\newcommand{\NHarnessUsable}{14}
\newcommand{\PctHarnessUsable}{7.5}
\newcommand{\NFailUndecl}{86}
\newcommand{\PctFailUndecl}{49.7}
\newcommand{\NFailSig}{29}

\newcommand{\NFailNoMem}{28}

\newcommand{\NFailLink}{14}

\newcommand{\NFailIncl}{7}

\newcommand{\NRuleExpr}{12}
\newcommand{\NNonExpr}{6}
\newcommand{\NToolMapped}{956}
\newcommand{\NToolClasses}{14}
\newcommand{\PctToolExpr}{96.8}
\newcommand{\NToolNonExpr}{31}
\newcommand{\NLLMMapped}{1{,}431}

\newcommand{\NLLMNonExpr}{256}
\newcommand{\PctLLMNonExpr}{17.9}
\newcommand{\NToolNonExprClasses}{2}
\newcommand{\NToolNonExprOne}{22}

\newcommand{\NToolNonExprTwo}{9}

\newcommand{\NBestEffortSilent}{3}
\newcommand{\NClassLLMOnly}{4}
\newcommand{\NClassLLMOnlyNonExpr}{4}
\newcommand{\ClassLLMOnlyList}{002, 010, 013, 017}
\newcommand{\NClusters}{1{,}691}
\newcommand{\NClustersLLMOnly}{175}

\newcommand{\HwCPU}{Intel Core i9-14900F (32 threads)}
\newcommand{\HwGPU}{NVIDIA RTX 4090 (23~GiB)}
\newcommand{\HwRAM}{62~GiB}

\begin{document}


\title{LLMSec-AV: A Vulnerability Taxonomy and LLM-Driven Software Weakness Discovery Framework for Autonomous Vehicles}

\TRBauthor*{Md. Wasiul Haque}{Department of Civil, Construction \& Environmental Engineering, The University of Alabama}{mhaque16@crimson.ua.edu}[2009 Smart Communities and Innovation Building (SCIB), 28 Kirkbride Lane,\\
  Tuscaloosa, AL 35487-0288][0009-0007-8417-9261]
\TRBauthor{Sagar Dasgupta, Ph.D.}{Department of Civil, Construction \& Environmental Engineering, The University of Alabama}{sdasgupta@ua.edu}[2009 Smart Communities and Innovation Building (SCIB), 28 Kirkbride Lane,\\
  Tuscaloosa, AL 35487-0288][0000-0001-8491-662X]
\TRBauthor{Mizanur Rahman, Ph.D.}{Department of Civil, Construction \& Environmental Engineering, The University of Alabama}{mizan.rahman@ua.edu}[2007 Smart Communities and Innovation Building (SCIB), 28 Kirkbride Lane,\\
  Tuscaloosa, AL 35487-0288][0000-0003-1128-753X]

\AuthorHeaders{Haque, Dasgupta, and Rahman}


\maketitle

\section{Abstract}
\hfill\break%
\noindent\textbf{Objectives:}~Automated vehicles rely on millions of lines of safety-critical software, but general-purpose analysis tools do not understand which code can affect vehicle motion. This study asks whether providing large language models~(LLMs) with explicit automated vehicle~(AV) security knowledge helps them detect weaknesses that rule-based tools cannot represent.

\hfill\break%
\noindent\textbf{Methods:}~We developed an AV vulnerability taxonomy containing \NClasses{} weakness classes derived from vulnerability records, security advisories, and AV-security literature. We then created LLM-based Security Analysis for Automated Vehicles~(LLMSec-AV) and evaluated it on Autoware. The framework decomposed \NTU{} translation units into \NUnits{} functions and analyzed \NUnitsEval{} functions under four prompting conditions involving taxonomy context, retrieval from \NCorpusIndexed{} prior disclosures, and multi-step analysis. Findings were compared with \NGroundTruth{} weakness locations mined from upstream fixes and with a flag-volume-matched permutation baseline. CodeQL, Semgrep, cppcheck, and the Clang Static Analyzer evaluated the same code, with AV-specific rules added to CodeQL and Semgrep. Generated fuzzing harnesses were also tested using AFL\raisebox{0.15ex}{++} and sanitizers.

\hfill\break%
\noindent\textbf{Findings:}~The LLM conditions recovered up to \PctCodTaxRecall\% of the \NGroundTruth{} known weakness locations, substantially outperforming the conventional analyzers. CodeQL, Semgrep, and the Clang Static Analyzer matched none, while cppcheck matched one despite producing \NCppcheckFindings{} alerts. This advantage was not attributable to the taxonomy alone, because unaided prompting achieved similar detection performance. However, taxonomy context increased the proportion of findings assigned to a weakness class from near zero to more than four-fifths, substantially improving interpretability and triage. The analysis also showed that \NNonExpr{} of the \NClasses{} classes cannot be directly represented as static-analysis rules. 


\hfill\break%
\noindent\textbf{Novelty:}~This study introduces an AV-specific vulnerability taxonomy used directly as machine-readable input for weakness discovery,  evaluated on real AV code against known weaknesses.

\hfill\break%
\noindent\textbf{Practical Applications:}~LLMSec-AV can help developers and reviewers organize and prioritize safety-relevant findings alongside conventional analyzers.

\newpage

\section{Introduction}\label{sec:intro}
An automated driving system is, in operational terms, a software system. Perception, localization, prediction, planning, and control are implemented as communicating components, and every steering, acceleration, or braking command is the result of computation. The safety case for an autonomous vehicle (AV) therefore depends in part on the correctness of several million lines of code. Functional-safety practice recognizes software faults as a systematic source of hazardous behavior~\citep{iso26262}, while current regulatory and cybersecurity frameworks require manufacturers to establish management processes covering the software throughout the vehicle life cycle~\citep{unece155,iso21434}. Transportation agencies assessing automated-driving deployments, and developers preparing them, consequently need practical methods for identifying software weaknesses that can affect vehicle motion. Empirical studies show that such defects occur at scale in automated-driving codebases~\citep{garcia2020comprehensive,lou2022study}.

Existing static-analysis tools were not designed around this vehicle context. CodeQL~\citep{avgustinov2016ql}, Semgrep~\citep{semgrep}, cppcheck~\citep{cppcheck}, and the Clang Static Analyzer~\citep{clangsa} detect general programming weaknesses such as buffer overruns, null dereferences, and resource leaks. However, they do not inherently represent which data reach a steering actuator, which inputs are externally influenceable, or when a missing validation check constitutes a transportation-safety hazard rather than a routine implementation defect \cite{haque2027security}. In AV software, this distinction is critical. A missing finiteness check may be inconsequential in a logging utility but hazardous in a lateral controller, because not-a-number is a valid value for floating-point message fields~\citep{ieee754}. The severity of a weakness therefore depends not only on its code pattern, but also on its location, reachability, and vehicle-level consequence.

Large language models (LLMs) offer a possible way to incorporate this context. Their code-generation and reasoning capabilities are well established~\citep{chen2021codex,fan2023llmse}, and they have been applied to vulnerability detection with mixed results on generic corpora~\citep{zhou2019devign,chakraborty2021deep}. Yet an LLM prompted only to identify ``security bugs'' in AV source code has no explicit model of actuation reachability or safety impact \cite{haque2026llm}. Our premise is that effective AV-specific analysis requires structured domain knowledge that defines what constitutes an AV-specific weakness and why it matters.

We therefore adopt a taxonomy-first approach. We construct an AV vulnerability taxonomy comprising \NClasses{} classes derived from public disclosure records~\citep{nvd} and AV-security literature~\citep{dieber2017ros,mayoralvilches2022sros2,maggi2022dds}. Each class is operationally defined, mapped to a Common Weakness Enumeration~(CWE) identifier~\citep{cwe}, instantiated in ROS~2 idioms~\citep{macenski2022ros2}, and associated with its potential vehicle-level consequence. The taxonomy then serves as machine-readable context for LLMSec-AV, our analysis pipeline, allowing us to evaluate whether explicit AV-domain knowledge improves weakness discovery.

Two design commitments distinguish the study. All measurements are obtained from a real AV stack rather than a curated benchmark: Autoware~\citep{kato2018autoware}, analyzed at a pinned commit and through its native build configuration so that the static analyzer, language model, and fuzzer operate on the same code. We also distinguish \emph{flagged} from \emph{confirmed} findings. A weakness reported by either an LLM or a static analyzer remains a hypothesis until a fuzzing harness~\citep{fioraldi2020aflpp} reproduces a sanitizer-detected fault~\citep{serebryany2012asan} at the reported location. This paper addresses four research questions:

\emph{\textbf{RQ1:}} Does taxonomy-guided, retrieval-augmented analysis improve weakness discovery over unaided LLM prompting on AV source code?
    
\emph{\textbf{RQ2:}} How does LLMSec-AV compare with established general-purpose static analyzers on the same code?
    
\emph{\textbf{RQ3:}} Which AV weakness classes do rule-based analyzers structurally fail to express, and does domain-grounded analysis reach them?
    
\emph{\textbf{RQ4:}} Which component: taxonomy context, retrieval, or multi-step structuring accounts for the observed change in the vulnerability discovery outcome?

These questions cannot be answered using existing artifacts. The required AV-domain knowledge is not yet available in machine-readable form, and no annotated AV codebase provides confirmed weaknesses against which detection performance can be measured. Addressing the research questions therefore required constructing the taxonomy, analysis pipeline, and evaluation benchmark that the study itself evaluates. The paper makes the following contributions:

\begin{itemize}[leftmargin=*]

\item We construct a machine-readable AV vulnerability taxonomy comprising \NClasses{} weakness classes, including \NClassAV{} that are specific to the cyber-physical driving context. Each class is operationally defined, mapped to a CWE identifier~\citep{cwe}, instantiated in ROS~2 idioms, and programmatically linked to supporting disclosed vulnerabilities.

\item We develop LLMSec-AV, a pipeline that uses the taxonomy as structured analysis context, combines it with retrieval over prior disclosures, and performs multi-step weakness analysis.

\item We assemble an evaluation benchmark on a real AV stack in which every reported weakness is submitted for dynamic confirmation, allowing \emph{flagged} and \emph{confirmed} findings to be measured separately.

\item We compare LLMSec-AV with four general-purpose static analyzers configured for this domain, adding hand-written AV-specific rules to the two that accept them and running the others at their broadest setting, so the comparison is not against stock defaults.

\end{itemize}

The remainder of the paper is organized as follows. Section~\ref{sec:related} reviews AV platforms, vehicle and middleware security, and source-level weakness discovery. Section~\ref{sec:threat} defines the threat model and experimental scope. Section~\ref{sec:taxonomy} presents the vulnerability taxonomy, and Section~\ref{sec:method} describes the LLMSec-AV pipeline. Section~\ref{sec:setup} details the evaluation configuration and domain-adapted baselines. Section~\ref{sec:findings} reports the results, Section~\ref{sec:discussion} interprets them, and Section~\ref{sec:conclusions} summarizes the limitations and future directions.

\section{Related Work}\label{sec:related}

This work draws on automated-driving platforms, vehicle security, and source-level weakness discovery. Together, these areas reveal a gap: prior research has characterized network, middleware, and sensor threats extensively, but offers limited support for detecting application-level weaknesses whose significance depends on automated-driving context.

\subsection{Automated Driving Software Platforms}

Proprietary platforms such as Waymo, Zoox, Aurora, Mobileye, and Tesla are not available for independent source-level analysis, so their safety claims reach reviewers as process documentation, assessment reports, and operational statistics rather than as inspectable artifacts. Levels of automation follow SAE~J3016~\citep{saej3016}. Among open stacks, Autoware~\citep{kato2018autoware,autoware} provides a full ROS~2-based pipeline spanning sensing through control~\citep{macenski2022ros2}, Apollo~\citep{apollo} provides comparable scope through CyberRT, and openpilot driver assistance only. Because source-analysis methods require accessible code and reproducible builds, open stacks are the practical basis for this study.

\subsection{Security of Automated and Connected Vehicles}

Early studies demonstrated remote compromise of production in-vehicle networks~\citep{koscher2010experimental,checkoway2011comprehensive}. Later work examined the ROS trust model~\citep{dieber2017ros,white2019sros}, optional DDS security in SROS2~\citep{mayoralvilches2022sros2}, and the DDS/RTPS transport attack surface~\citep{maggi2022dds}. Sensor spoofing has also been studied extensively~\citep{cao2019adversarial,shin2017illusion}, but is outside our scope. Empirical analyses of AV software report recurring numeric, validation, and error-handling defects~\citep{garcia2020comprehensive,lou2022study}. Unlike these descriptive studies, our taxonomy defines weakness classes through an explicit attacker model, CWE mappings, ROS~2 manifestations, and vehicle-level consequences, then uses them as analysis context. RVSS informs the treatment of severity~\citep{vilches2018rvss}.

\subsection{Weakness Discovery in Source Code}

The evaluated analyzers represent complementary strategies. CodeQL supports declarative interprocedural analysis~\citep{avgustinov2016ql}; Semgrep performs syntactic matching with limited dataflow~\citep{semgrep}; cppcheck applies broad local checks~\citep{cppcheck}; and the Clang Static Analyzer performs path-sensitive symbolic execution within translation units~\citep{clangsa}. None natively models ROS~2 message sources, actuation reachability, or vehicle-level consequences. Learning-based vulnerability detection has focused mainly on generic CWE-oriented C/C\raisebox{0.15ex}{++} and Java datasets~\citep{zhou2019devign}, with results sensitive to dataset construction~\citep{chakraborty2021deep}. That literature leaves open whether models can detect weaknesses whose meaning depends on deployment context. LLMSec-AV draws on chain-of-thought prompting~\citep{wei2022cot}, retrieval-augmented generation~\citep{lewis2020rag}, and model-assisted fuzz-harness generation~\citep{xia2024fuzz4all}. Our ablation evaluates retrieval over disclosed vulnerabilities, while the confirmation stage examines whether generated harnesses can validate findings in a production ROS~2 stack.

\section{Threat Model and Scope}\label{sec:threat}
This section defines the adversary capabilities, middleware (ROS~2) execution properties, and evaluation boundary used throughout the study. The threat model follows ROS~2's default trust posture: unless optional security plugins are provisioned, participants with DDS transport access may publish to topics, invoke services, or modify parameters~\citep{mayoralvilches2022sros2,omgdds}.

\subsection{Terminology}

Following the CWE catalogue~\citep{cwe}, a \emph{weakness} is a defect class, whereas a \emph{vulnerability} is a specific exploitable instance. Section~\ref{sec:taxonomy} defines weakness classes. A \textbf{flagged} finding is a location reported by an LLM or analyzer; a \textbf{confirmed} finding additionally produces an attributable sanitizer-detected fault within the fuzzing budget. Unconfirmed findings remain unresolved rather than being treated automatically as false positives.

\subsection{Experimental Case}

Autoware~\citep{kato2018autoware} serves as a reproducible case study because it is full-stack, production-representative, buildable at a pinned revision with a compilation database, and accompanied by public development history for constructing the ground truth in Section~\ref{sec:groundtruth}. Apollo~\citep{apollo} is a natural second case but uses CyberRT rather than ROS~2, while openpilot covers a narrower driver-assistance scope. The single-case design supports analytic rather than statistical generalization, and transfer to another stack would require adapting the taxonomy's middleware-specific instantiations, as discussed in Section~\ref{sec:conclusions}.

Autoware consists of ROS~2 nodes exchanging typed DDS messages across sensing, perception, localization, planning, control, and actuation~\citep{omgdds,macenski2022ros2}. Three properties shape the threat surface. First, message typing does not ensure semantic validity: \texttt{float64} fields may contain NaN or infinity under IEEE~754~\citep{ieee754}. Second, concurrency depends on callback groups and executor configuration rather than explicit thread creation~\citep{macenski2022ros2,ros2design}. Third, subscribers cannot distinguish unchanged data from publisher failure without timestamp or DDS-liveliness checks~\citep{omgdds}. Malformed, delayed, or absent messages may therefore affect vehicle behavior.

\subsection{Attacker Capabilities}

The adversary may publish to reachable topics, invoke services, modify parameters, delay or suppress streams, and supply off-vehicle inputs forwarded internally, including maps, calibration files, V2X messages, and teleoperation commands~\citep{mayoralvilches2022sros2}. Physical sensor spoofing, hardware attacks, side channels, and supply-chain compromise are outside scope.

\subsection{Evaluation Boundary}

The target is the C\raisebox{0.15ex}{++} portion of Autoware on ROS~2 Humble. The analyzable subset contains \NTU{} non-test, non-generated translation units represented in the workspace compilation database. Reusing their native compilation commands ensures consistent type resolution across parsing, CodeQL extraction, and harness compilation for the comparison in Section~\ref{sec:findings}.

\section{AV Vulnerability Taxonomy}\label{sec:taxonomy}

In automated-driving software, weakness significance depends on the affected function, input path, middleware context, and vehicle-level consequence. The taxonomy therefore includes only classes that are specific to or materially amplified by automated driving and grounds each in disclosed vulnerabilities, published research, or documented middleware behavior.

\subsection{Construction}
The taxonomy draws from three sources: NVD records and public advisories, AV and robotics security literature, and documented ROS~2 and DDS defaults. NVD records were retrieved through CVE API queries covering robotics, middleware, automotive, and embedded-control terms~\citep{nvd}. Exact-phrase queries and domain-anchor filtering reduced off-domain matches. Of \NCVERetrieved{} retrieved entries, \NCVERetained{} were retained; rejected entries are preserved for audit.

\subsection{Structure and Evidence}
\begin{figure}[ht]
  \centering
  \includegraphics[width=\linewidth]{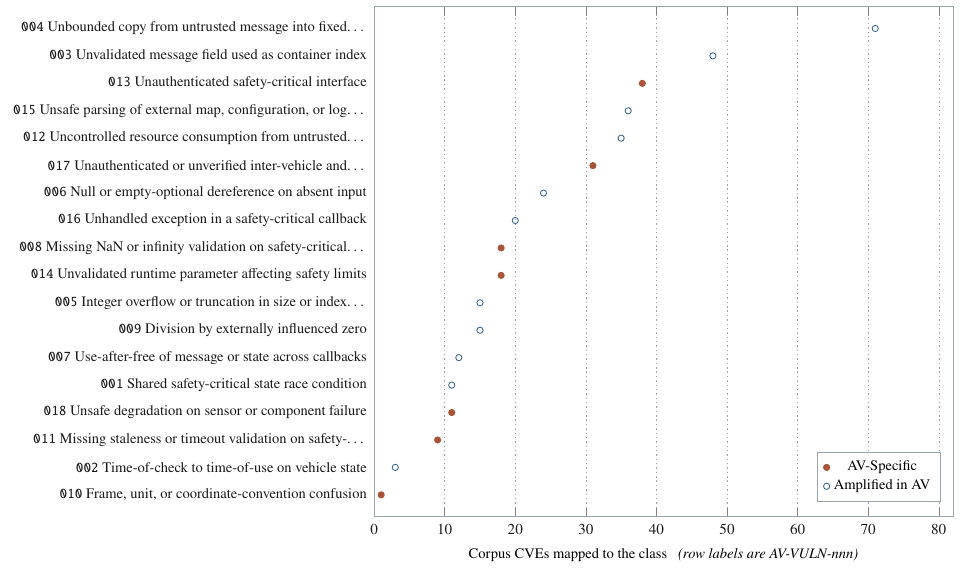}
  \caption{Supporting evidence per taxonomy class, ordered by the number of
  corpus vulnerability records whose CWE mapping intersects the class. Filled
  markers denote AV-specific classes; hollow markers denote classes amplified in
  an AV context.}
  \label{fig:taxonomy}
\end{figure}

The taxonomy contains \NClasses{} classes, \NClassAV{} AV-specific and \NClassAmp{} AV-amplified, spanning \NDistinctCWE{} primary CWE identifiers. Each records its CWE mapping, provenance, code pattern, ROS~2 instantiation, vehicle-level consequence, detection notes, severity prior, and ISO/SAE~21434 impact rating on the four-level scale~\citep{iso21434}, assuming SAE Level~4 operation without an attentive safety driver. CWE intersection links the classes to \NEvidenceLinks{} supporting records. Figure~\ref{fig:taxonomy} shows that AV-specific classes carry less disclosure evidence than amplified ones, reflecting the limits of public vulnerability records discussed in Section~\ref{sec:conclusions}. Severity priors and impact ratings are retained in the artifact but withheld from the model context (Section~\ref{sec:setup}).

\begin{table}[tb]
  \caption{The AV vulnerability taxonomy. \emph{Scope} distinguishes classes with no meaning outside a driving context (AV) from generic weaknesses whose consequence class changes in a vehicle (Amp). \emph{Ev.} is the number of corpus vulnerability records whose CWE mapping intersects the class. \emph{Rule} indicates whether the class is expressible as a static-analysis rule.}
  \label{tab:taxonomy}
  \centering\footnotesize
  \begin{tabularx}{\linewidth}{@{}l Xl cc r@{}}
    \toprule
    ID & Weakness class & Primary CWE & Scope & Rule & Ev. \\
    \midrule
    001 & Shared safety-critical state race condition & CWE-362 & Amp & \checkmark & 11 \\
    002 & Time-of-check to time-of-use on vehicle state & CWE-367 & Amp & --- & 3 \\
    003 & Unvalidated message field used as container index & CWE-129 & Amp & \checkmark & 48 \\
    004 & Unbounded copy from untrusted message into fixed buffer & CWE-120 & Amp & \checkmark & 71 \\
    005 & Integer overflow or truncation in size or index arithmetic & CWE-190 & Amp & \checkmark & 15 \\
    006 & Null or empty-optional dereference on absent input & CWE-476 & Amp & \checkmark & 24 \\
    007 & Use-after-free of message or state across callbacks & CWE-416 & Amp & \checkmark & 12 \\
    008 & Missing NaN or infinity validation on safety-critical value & CWE-1077 & AV & \checkmark & 18 \\
    009 & Division by externally influenced zero & CWE-369 & Amp & \checkmark & 15 \\
    010 & Frame, unit, or coordinate-convention confusion & CWE-682 & AV & --- & 1 \\
    011 & Missing staleness or timeout validation on safety-critical input & CWE-672 & AV & --- & 9 \\
    012 & Uncontrolled resource consumption from untrusted message & CWE-400 & Amp & \checkmark & 35 \\
    013 & Unauthenticated safety-critical interface & CWE-306 & AV & --- & 38 \\
    014 & Unvalidated runtime parameter affecting safety limits & CWE-1284 & AV & \checkmark & 18 \\
    015 & Unsafe parsing of external map, configuration, or log input & CWE-502 & Amp & \checkmark & 36 \\
    016 & Unhandled exception in a safety-critical callback & CWE-248 & Amp & \checkmark & 20 \\
    017 & Unauthenticated or unverified inter-vehicle and remote input & CWE-345 & AV & --- & 31 \\
    018 & Unsafe degradation on sensor or component failure & CWE-754 & AV & --- & 11 \\
    \bottomrule
  \end{tabularx}
\end{table}

\subsection{Example Classes in Detail}
\textbf{AV-VULN-001, shared safety-critical state race condition.} A member
variable holding vehicle state (pose, velocity, trajectory, gate mode, emergency
flag) is written by one callback and read by another without mutual exclusion:

\begin{lstlisting}
// subscription callback (executor thread A)
void onOdom(const Odometry::ConstSharedPtr msg) {
  current_pose_ = msg->pose.pose;             // unguarded write
}
// timer callback (executor thread B, reentrant group)
void onTimer() {
  auto cmd = computeControl(current_pose_);   // unguarded read: may tear
  pub_->publish(cmd);
}
\end{lstlisting}

A torn pose or velocity read may produce a control command for a state the
vehicle never occupied, yet remain indistinguishable from a valid command, and an
adversary controlling publication rate on the topic can influence callback
timing. Because the race arises from ROS~2 executor and callback-group semantics
rather than explicit thread creation, analyzers modeling only standard threading
and locking constructs may miss it.

\textbf{AV-VULN-008, missing finiteness validation.} This class has no
syntactic defect at all:

\begin{lstlisting}
const double err = target_x_ - msg->pose.pose.position.x;  // may be NaN
double steer = kp_ * err;
if (steer >  max_steer_) steer =  max_steer_;  // NaN passes: NaN > x is false
if (steer < -max_steer_) steer = -max_steer_;  // NaN passes as well
publishSteering(steer);                        // NaN reaches the actuator
\end{lstlisting}

The code is syntactically valid, but NaN bypasses both comparisons and reaches the actuator. Detection requires combining IEEE~754 semantics, legal NaN values in ROS~2 \texttt{float64} fields, and the actuator-facing sink. Similar context-dependent reasoning is required for AV-VULN-010, AV-VULN-011, and AV-VULN-018, making these classes central to RQ3.

\subsection{Transferability}
Class definitions are middleware-neutral, while separate instantiation fields encode ROS~2 idioms. Retargeting should therefore require revising the \NClasses{} instantiations rather than reconstructing the taxonomy. This remains a design objective because the study evaluates only one stack.


\section{LLMSec-AV Pipeline}\label{sec:method}
\begin{figure}[!tb]
  \centering
  \includegraphics[width=\linewidth]{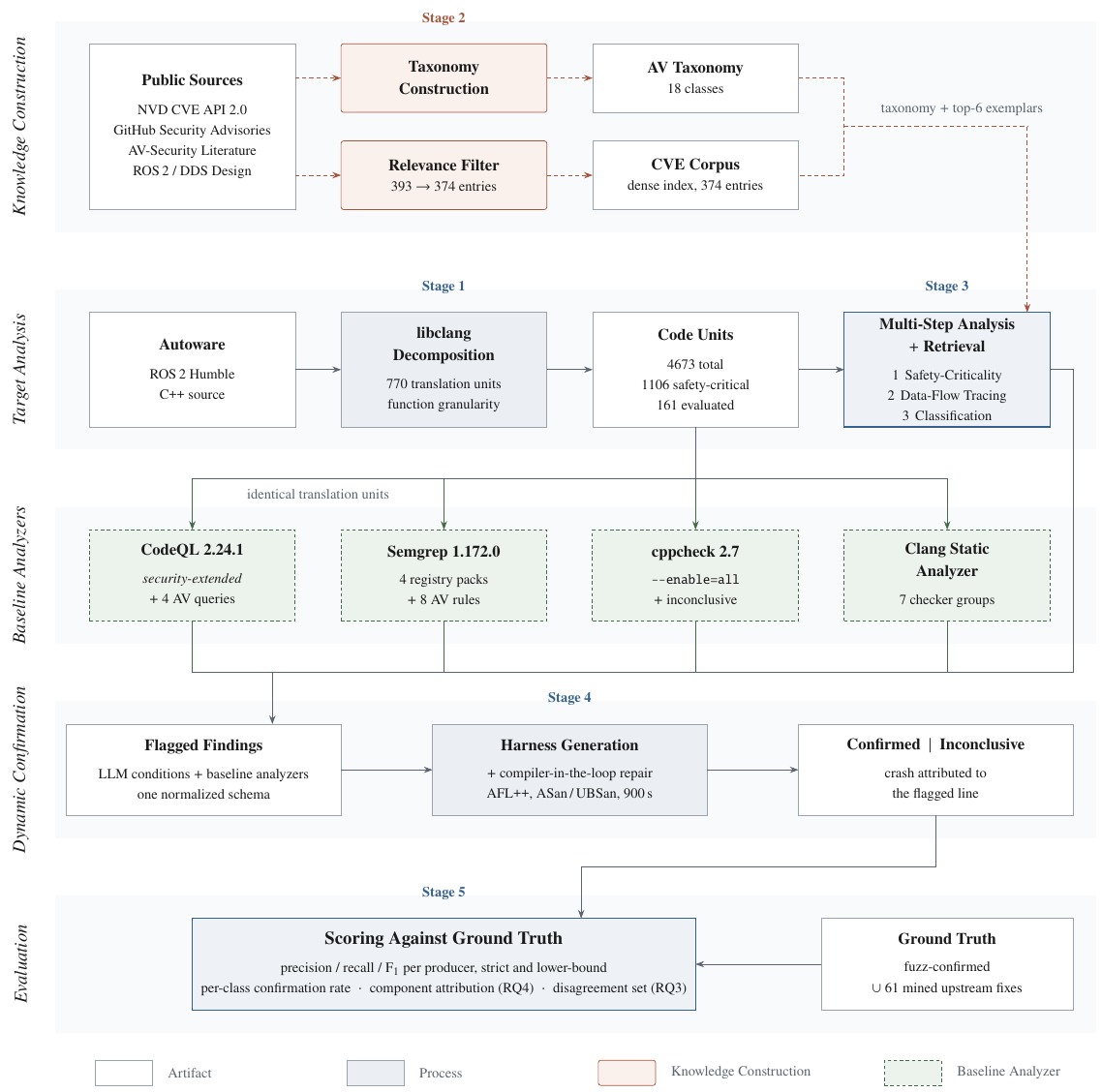}
  \caption{The LLMSec-AV pipeline. Knowledge construction produces the taxonomy
  and the retrieval corpus; target analysis decomposes the stack and analyzes
  each code unit; the four baseline analyzers consume the identical translation
  units; and flagged findings from both families are submitted for dynamic
  confirmation on the same terms before scoring.}
  \label{fig:pipeline}
\end{figure}

LLMSec-AV analyzes the target stack at function level. It identifies functions that may influence vehicle motion, supplies each to the language model together with the taxonomy and the most relevant prior disclosures, and submits flagged findings for dynamic confirmation against the compiled implementation. The four general-purpose analyzers evaluate the same units, and findings from all producers are assessed against a common ground truth. Each comparison varies only one factor: all four prompting conditions use the same task instruction and output contract, so wording differences cannot confound the results, and dynamic confirmation is applied uniformly to every producer, since constructing ground truth from one producer's candidates would merely reproduce that producer's output rather than evaluate it. Figure~\ref{fig:pipeline} summarizes the five implementation stages.

\subsection{Stage 1: Decomposition and Safety-Critical Tagging}

Each of the \NTU{} translation units is parsed with its native compilation flags, producing \NUnits{} function-level code units with exact source extents; \NParseFail{} unit failed to parse. Function-level granularity aligns with the line-level output of the four baseline analyzers, fits within the model context window, and supports targeted fuzz-harness generation. Cross-function behavior is represented through the enclosing-file context supplied with each unit.

Safety-critical tagging proceeds in two steps. A deterministic keyword filter spanning six signal groups (actuation, perception fusion, planning decision, localization, concurrency and shared state, and untrusted entry) identifies candidates and records the supporting evidence. It tags \NUnitsSC{} of \NUnits{} units (\PctSC\%). The first model-analysis step then confirms or rejects each tag, allowing tagging agreement to be measured.

To control analysis cost, we use a seeded stratified sample selected round-robin across pipeline-domain and primary-signal cells. This yields \NUnitsStrat{} units covering all domains and signal groups. We add \NUnitsKnown{} units containing mined real-world issues from Section~\ref{sec:groundtruth}, producing \NUnitsEval{} evaluated units. This augmentation is independent of producer performance, precedes all analysis, and is applied identically to every producer. The stratified and known-issue strata are retained and reported separately.

\subsection{Stage 2: Taxonomy and Retrieval Corpus}

The taxonomy is described in Section~\ref{sec:taxonomy}. The retrieval corpus contains \NCorpusIndexed{} entries, comprising \NCorpusNVD{} vulnerability records and \NCorpusAdv{} advisories, embedded with \EmbedModel{} into \EmbedDim{}-dimensional vectors. Retrieval uses exhaustive cosine search to preserve recall and reproducibility. For each unit, the system returns the top \RagTopK{} entries above a similarity threshold of \RagMinSim{}. Queries are constructed from the unit's identity and body rather than from taxonomy labels, avoiding circularity in the RQ4 ablation.

\subsection{Stage 3: Structured Multi-Step Analysis}

The proposed condition uses three model invocations, with each structured output passed verbatim to the next:

\begin{enumerate}[leftmargin=*]
\item \textbf{Safety-criticality:} determine whether the unit is safety-critical and identify the path by which its outputs may influence actuation.
\item \textbf{Data-flow tracing:} identify externally influenced sources, propagation paths, existing guards, and unguarded flows to dangerous operations.
\item \textbf{Classification:} match the traced hazardous pattern to the taxonomy using retrieved disclosures and emit a structured finding grounded in the preceding trace.
\end{enumerate}

\begin{figure}[!tb]
  \centering
  \includegraphics[width=\linewidth]{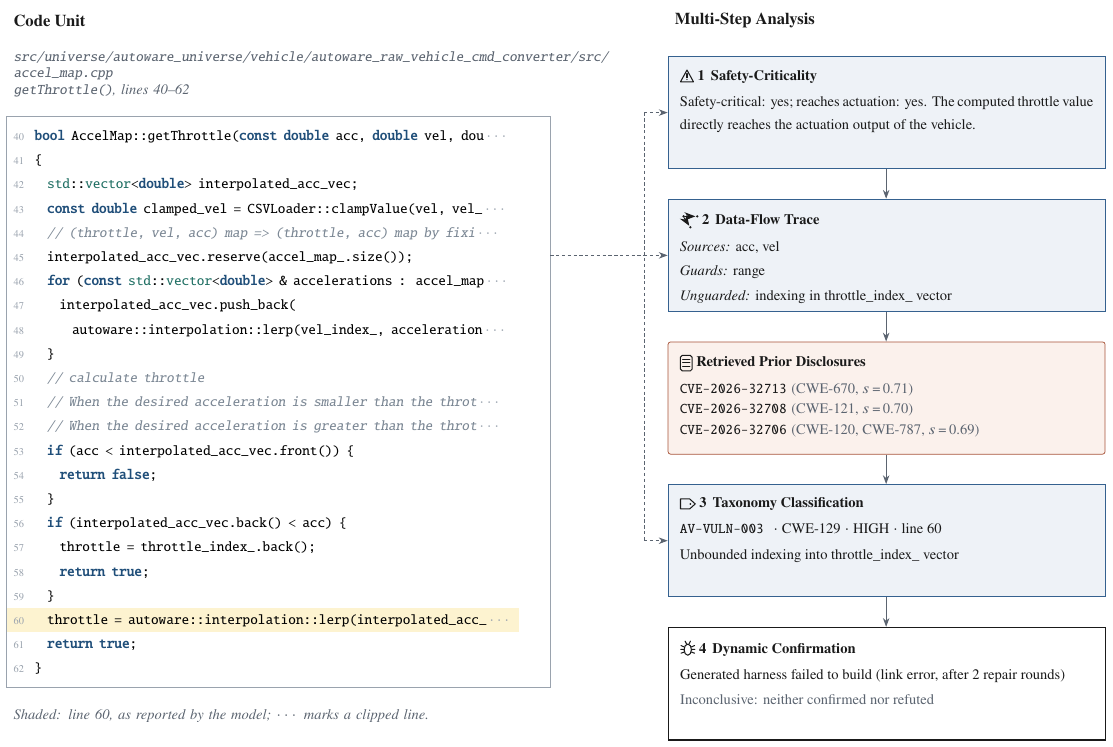}
  \caption{One code unit carried end to end under the \texttt{full} condition. The source
  under analysis is supplied to all three analysis steps; retrieved prior disclosures enter at the
  classification step; and the structured finding is then submitted for dynamic
  confirmation.}
  \label{fig:walkthrough}
\end{figure}

We call this \emph{structured multi-step analysis} rather than model-internal reasoning because the intermediate trace is explicitly carried between prompts. Step~3 therefore conditions on a concrete artifact naming sources, guards, and unguarded flows rather than reconstructing them internally. This design does not require a reasoning-tuned model. Figure~\ref{fig:walkthrough} illustrates one finding from initial analysis through final classification.

\subsection{Ablation Design}
Four conditions isolate each component's contribution. \texttt{zero\_shot} is unaided prompting; \texttt{taxonomy\_only} adds the taxonomy;
\texttt{taxonomy\_rag} adds retrieved disclosures; and \texttt{full} adds the
three-step decomposition. The task instruction and output contract are
byte-identical across all four; only the injected context differs and, for the
full condition, the decomposition into three invocations. Reported line numbers
are the file's real line numbers, so a finding maps onto ground truth and onto
baseline output with no translation step that could misalign the comparison.


\subsection{Stage 4: Dynamic Confirmation}

Each flagged finding is assigned a generated harness that reads fuzzer input from standard input and preserves hostile values such as NaN, infinity, zero, negatives, and extreme magnitudes. It is compiled with the target translation unit using that unit's native build flags under AFL\raisebox{0.15ex}{++}~\citep{fioraldi2020aflpp} with AddressSanitizer~\citep{serebryany2012asan} and UndefinedBehaviorSanitizer, then linked against the installed package libraries.

Build failures are classified using a fixed taxonomy and retried for up to \NRepairRounds{} compiler-in-the-loop rounds, with compiler diagnostics returned to the model, identically across conditions. Each crash is replayed under the sanitizers and attributed only if the stack contains a frame in the flagged file within \MatchTol{} lines of the reported unit. Outcomes are \emph{confirmed}, \emph{crash unattributed}, \emph{no crash in budget}, \emph{harness build failed}, or \emph{harness generation failed}, using a budget of \FuzzBudget{}~seconds per finding. The design admits baseline findings on the same terms, since otherwise a baseline finding at a location no model flagged could never enter the ground truth and baseline recall would be scored against a ground truth the baseline was barred from contributing to. Under the compute available, confirmation was in fact run over \NLLMSubmitted{} model findings only, with \NBaselineSubmitted{} baseline findings submitted; the pipeline records this as a pool-bias flag, whose consequence Section~\ref{sec:findings} shows to be narrower than the flag implies.

\subsection{Ground Truth}\label{sec:groundtruth}

Ground truth combines two independent sources. The first consists of locations confirmed through fuzzing in Stage~4. The second uses security-relevant fixes committed after the pinned Autoware revision. Because public Autoware vulnerability records rarely identify files and lines, each qualifying fix is treated as evidence that the corresponding pre-fix code was weak, and its diff is used to localize the affected lines. We examined \NCommitsScanned{} newer upstream commits across repositories containing evaluation targets. The selection rule requires either a fix-class commit or an explicit vulnerability reference. Commits producing more than ten candidate locations are excluded as bulk refactors; this removed \NBulkRejected{} commits containing \NBulkRejectedLocs{} locations. The final set contains \NKnownIssues{} locations: \NKnownHigh{} high-confidence cases in which the fix adds a recognized guard adjacent to the affected line, and \NKnownMed{} medium-confidence cases. Merging locations that fall within the \MatchTol{}-line matching tolerance of one another, so that two adjacent pre-fix lines count as one weakness rather than inflating every recall denominator, leaves the \NGroundTruth{} distinct locations scored in Section~\ref{sec:findings}. Because this mapping remains heuristic, its implications are discussed in Section~\ref{sec:conclusions}.

\subsection{Stage 5: Scoring}

A finding is a true positive if it falls within \MatchTol{} lines of a ground-truth location in the same file. Ground-truth locations within this tolerance are merged so that adjacent reports count as one weakness rather than inflating recall.

Every unconfirmed finding counts as a false positive, so reported precision is a lower bound whose denominator is the producer's flag volume. A variant that excludes inconclusive findings from both numerator and denominator is computed in the released artifacts; because Stage~4 decided almost nothing, the two agree closely and only the former is reported here. Findings are normalized into a common category space using, in order, an explicit taxonomy identifier, a custom-rule self-tag, a rule-to-category mapping, or the reported CWE. Unmapped findings are retained and counted, since the unmapped fraction itself reflects category coverage. Recall alone cannot separate localization from flag volume. A producer reporting densely in the units that hold the ground-truth locations will match some by coincidence, because a match requires only landing within \MatchTol{} lines of the true line, and the producers compared here differ in volume by more than an order of magnitude. We therefore score every recall figure against a permutation null that holds each producer's flag count and per-file distribution fixed and destroys only its within-file line information, resampling each finding's line uniformly from the span of the code units that producer reported in that file. The resulting null recall is what volume alone would yield, and the difference from observed recall is the component attributable to localization. Because the control is constructed per producer, it is fair between producers of very different volume: whichever saturates a file is granted the same saturation in its own null. We draw \NullIters{} permutations, placing the one-sided resolution floor at $p\le\NullPFloor{}$.

\section{Evaluation Methodology and Configuration}\label{sec:setup}

The evaluation uses a pinned Autoware revision (\AwCommit) on ROS~2 Humble and analyzes \NUnitsEval{} code units, \NUnitsStrat{} drawn by stratified sampling from the \NUnitsSC{} safety-critical functions and \NUnitsKnown{} added because they contain a mined real-world issue, out of \NUnits{} functions across \NTU{} compilable translation units. All tools use the target units' native compilation settings. Model inference uses temperature 0.1, top-$p$ 0.9, and a fixed seed. Retrieval indexes \NCorpusIndexed{} disclosures with \EmbedModel{} embeddings and exact cosine search. Dynamic confirmation uses AFL\raisebox{0.15ex}{++} \AFLVer{} with AddressSanitizer and UndefinedBehaviorSanitizer for \FuzzBudget{}~seconds per finding. Findings are matched using a \MatchTol{}-line tolerance, with strict and lower-bound scoring.

We evaluate two locally hosted open-weight models with contrasting capabilities. Codestral 22B~\citep{codestral} is code-specialized but not reasoning-tuned, whereas gpt-oss 20B~\citep{gptoss} is reasoning-tuned. Because the multi-step condition passes intermediate artifacts explicitly between prompts, it does not depend on internal reasoning. The comparison therefore tests whether explicit decomposition remains useful across model types.

\subsection{Baseline Configuration}

RQ2 compares LLMSec-AV with CodeQL, Semgrep, cppcheck, and the Clang Static Analyzer, each configured for this domain rather than left at defaults. CodeQL uses \CodeQLPack{} with \NCustomQL{} AV-specific queries and Semgrep four registry packs with \NCustomSemgrep{} AV rules; custom and stock findings are reported separately. cppcheck and the Clang Static Analyzer accept no custom dataflow rules of this kind, so they are instead run at maximum breadth, with \texttt{-{}-enable=all} and inconclusive checks and with seven checker groups respectively. The two families are therefore not equally adaptable, and weaker results for the latter two on cross-callback classes are expected rather than a claim about them. The CodeQL extensions model ROS~2 subscription callback inputs as untrusted sources. Custom rules cover 12 of the \NClasses{} taxonomy classes, including deliberate best-effort rules for four classes that Table~\ref{tab:taxonomy} marks as not expressible in principle, so that the comparison is not weakened by our declining to try. The \emph{Rule} column therefore records expressibility rather than whether we wrote a rule. Section~\ref{sec:findings} reports what those best-effort rules produced. A CodeQL database already present in the target tree covered only nine of the evaluated units, having been built from an incomplete build; using it would have shown CodeQL finding almost nothing for reasons unrelated to its capability. We therefore rebuilt it by replaying the workspace's native compilation commands under the extractor, obtaining coverage of all \DBCovUnits{} evaluated units across all \DBCovFiles{} of their source files. Coverage is verified programmatically to avoid confounding analyzer performance with incomplete extraction.

All experiments ran on \HwCPU{}, \HwRAM{}, and \HwGPU{}, with complete software versions and configuration files included in the released artifacts.

\section{Findings}\label{sec:findings}
Three results dominate. First, all primary model conditions localized weaknesses later fixed upstream at rates exceeding those expected from their flag volumes, whereas none of the four static analyzers did, including the two carrying hand-written AV-specific rules. Second, this advantage belongs to model-based analysis rather than specifically to the taxonomy: unaided prompting also exceeded the null, while the taxonomy provided a modest recall gain and substantially improved class labeling. Third, dynamic confirmation produced no confirmed findings because most generated harnesses failed to integrate with the ROS~2 type and initialization surface.

\subsection{Analysis Inventory}
Across two models and four conditions, plus a medium-reasoning-effort replication of gpt-oss on a \NOssMedFullUnits{}-unit subset, we executed \NRunsTotal{} runs over \NUnitsEval{} code units. Table~\ref{tab:inventory} reports finding volume, taxonomy mapping, prompt truncation, and answer exhaustion. The latter two are validity failures rather than negative findings because the intended treatment was not completed.

\begin{table}[tb]
  \caption{Flagged-finding inventory by model and condition. \emph{Mapped} is the percentage of findings the producer assigned to a taxonomy class. \emph{Trunc.} counts runs whose prompt met the context ceiling and \emph{Exh.} counts runs that consumed the generation budget without answering; both are validity diagnostics and should be zero.}
  \label{tab:inventory}
  \centering\small
  \begin{tabular}{@{}llrrrrrr@{}}
    \toprule
    Model & Condition & Units & Findings & /unit & Mapped & Trunc. & Exh. \\
    \midrule
    Codestral 22B & \texttt{zero\_shot} & 161 & 259 & 1.61 & 0\% & 0 & 0 \\
     & \texttt{taxonomy\_only} & 161 & 351 & 2.18 & 82\% & 0 & 0 \\
     & \texttt{taxonomy\_rag} & 161 & 332 & 2.06 & 73\% & 0 & 0 \\
     & \texttt{full} & 161 & 237 & 1.47 & 81\% & 0 & 0 \\
    \midrule
    gpt-oss 20B & \texttt{zero\_shot} & 161 & 89 & 0.55 & 2\% & 0 & 0 \\
     & \texttt{taxonomy\_only} & 161 & 97 & 0.60 & 96\% & 0 & 0 \\
     & \texttt{taxonomy\_rag} & 161 & 101 & 0.63 & 91\% & 0 & 0 \\
     & \texttt{full} & 161 & 135 & 0.84 & 73\% & 0 & 0 \\
    \midrule
    gpt-oss 20B (medium) & \texttt{taxonomy\_rag} & 15 & 12 & 0.80 & 92\% & 0 & 0 \\
     & \texttt{full} & 15 & 18 & 1.20 & 100\% & 0 & 0 \\
    \bottomrule
  \end{tabular}
\end{table}

\subsection{Aggregate Comparison and the Volume-Matched Null (RQ1, RQ2)}
\label{sec:null}
Because Stage~4 confirmed no findings, ground truth consists solely of the \NGroundTruth{} deduplicated locations mined from upstream fixes. The resulting comparison is independent of which producers supplied candidates for confirmation.

Table~\ref{tab:comparison} shows a clear recall difference. The model conditions recovered between \NullOssZsObs{} and \NullCodTaxObs{} of the mined locations. cppcheck matched one location despite reporting \NCppcheckFindings{} findings, while CodeQL, Semgrep, and the Clang Static Analyzer matched none, including CodeQL and Semgrep under their AV-specific rule sets. This was not caused by missing coverage: the tools analyzed the relevant translation units and produced \NBaseInUnits{} of their \NBaseAll{} findings within the evaluated units. Flag volume explains part, but not all, of the model recall. All \NLLMConditions{} primary conditions exceeded their volume-matched nulls by \MinLLMLift{} to \MaxLLMLift{}, whereas none of the \NToolProducers{} analyzer configurations exceeded its null. The medium-effort replication rows did not exceed their nulls, but their small sample and finding counts provide insufficient power for interpretation. Zero-shot Codestral achieved \NullCodZsObs{} against a null of \NullCodZsExp{}, comparable to the taxonomy-guided conditions. Thus, RQ1 supports only a modest taxonomy-related recall gain, from \NullCodZsObs{} to \NullCodTaxObs{} for Codestral and from \NullOssZsObs{} to \NullOssTaxObs{} for gpt-oss, together with a substantial gain in categorization. For RQ2, the model conditions localized mined weaknesses at rates not explained by volume, while the rule-based analyzers did not.

Reported precision is only a lower bound. Ground truth identifies known weak locations but does not establish that findings elsewhere are false, so precision is bounded below by construction and tends to fall as flag volume rises. Estimating practical precision requires blind adjudication of findings at previously unknown locations.

\begin{table}[tb]
  \caption{Per-producer scores against 46 ground-truth locations, with the flag-volume-matched permutation null. \emph{P} and \emph{R} count an unconfirmed finding as a false positive, so \emph{P} is a lower bound whose denominator is flag volume. \emph{Null R} is the recall the producer would obtain from its flag volume alone over 2000 permutations, and \emph{Lift} is the remainder, attributable to localization. $p$ is one-sided with a resolution floor of $\le$0.0005.}
  \label{tab:comparison}
  \centering\small
  \begin{tabular}{@{}lrrrrrrr@{}}
    \toprule
    \multirow{2}{*}{Producer} & \multirow{2}{*}{$n$} & \multicolumn{3}{c}{Observed} & \multicolumn{3}{c}{Volume-matched null} \\
    \cmidrule(lr){3-5}\cmidrule(lr){6-8}
    & & P & R & F$_1$ & Null R & Lift & $p$ \\
    \midrule
    \multicolumn{8}{@{}l}{\emph{General-purpose static analyzers}} \\
    \quad Clang Static Analyzer & 65 & 0.000 & 0.000 & 0.000 & 0.008 & -0.008 & 1.000 \\
    \quad CodeQL (custom AV queries) & 692 & 0.000 & 0.000 & 0.000 & 0.061 & -0.061 & 1.000 \\
    \quad CodeQL (\texttt{security-extended}) & 111 & 0.000 & 0.000 & 0.000 & 0.006 & -0.006 & 1.000 \\
    \quad cppcheck & 1301 & 0.001 & 0.022 & 0.002 & 0.051 & -0.029 & 0.909 \\
    \quad Semgrep (custom AV rules) & 131 & 0.000 & 0.000 & 0.000 & 0.002 & -0.002 & 1.000 \\
    \midrule
    \multicolumn{8}{@{}l}{\emph{LLMSec-AV conditions}} \\
    \quad Codestral 22B \texttt{full} & 237 & 0.173 & 0.674 & 0.275 & 0.295 & +0.379 & $\le$0.0005 \\
    \quad Codestral 22B \texttt{taxonomy\_only} & 351 & 0.145 & 0.761 & 0.244 & 0.384 & +0.377 & $\le$0.0005 \\
    \quad Codestral 22B \texttt{taxonomy\_rag} & 332 & 0.123 & 0.652 & 0.208 & 0.353 & +0.299 & $\le$0.0005 \\
    \quad Codestral 22B \texttt{zero\_shot} & 259 & 0.158 & 0.696 & 0.258 & 0.327 & +0.369 & $\le$0.0005 \\
    \quad gpt-oss 20B \texttt{full} & 135 & 0.163 & 0.370 & 0.226 & 0.149 & +0.221 & $\le$0.0005 \\
    \quad gpt-oss 20B \texttt{taxonomy\_only} & 97 & 0.186 & 0.435 & 0.260 & 0.137 & +0.298 & $\le$0.0005 \\
    \quad gpt-oss 20B \texttt{taxonomy\_rag} & 101 & 0.178 & 0.326 & 0.231 & 0.145 & +0.181 & $\le$0.0005 \\
    \quad gpt-oss 20B \texttt{zero\_shot} & 89 & 0.157 & 0.217 & 0.182 & 0.075 & +0.142 & $\le$0.0005 \\
    \quad gpt-oss 20B (medium) \texttt{full} & 18 & 0.167 & 0.043 & 0.069 & 0.027 & +0.016 & 0.353 \\
    \quad gpt-oss 20B (medium) \texttt{taxonomy\_rag} & 12 & 0.333 & 0.065 & 0.109 & 0.041 & +0.024 & 0.212 \\
    \bottomrule
  \end{tabular}
\end{table}

\subsection{Component Attribution (RQ4)}
The ablation does not support a reliable component ranking. Table~\ref{tab:ablation} places F$_1$ between \AblRagF{} and \AblFullF{}, a spread of roughly four hundredths from one run per condition without variance estimates. The conditions nevertheless differ behaviorally. Taxonomy context increases recall from \AblZsR{} to \AblTaxR{} but also increases finding volume. Retrieval reduces both recall and volume. Multi-step structuring reduces volume while retaining recall near the taxonomy-only condition, suggesting greater selectivity rather than additional discovery. The clearest taxonomy effect is classification coverage: mapped findings rise from \PCodZsMapped\% to \PCodTaxMapped\% for Codestral and from \POssZsMapped\% to \POssTaxMapped\% for gpt-oss, as shown in Table~\ref{tab:inventory}.

Accordingly, RQ4 supports three limited conclusions: the taxonomy improves categorization and may modestly improve recall, retrieval did not help on this stack, and multi-step analysis trades volume for selectivity. None explains the broader localization advantage over static analyzers, which is already present under zero-shot prompting.

\begin{table}[tb]
  \caption{Component attribution (RQ4). Each row adds one component to the row above; $\Delta$F$_1$ is the change attributable to that component. Each condition pools the findings of both models, so its recall exceeds any single model's recall in Table~\ref{tab:comparison}.}
  \label{tab:ablation}
  \centering\small
  \begin{tabular}{@{}lcccrrrr@{}}
    \toprule
    Condition & Tax. & Ret. & Multi & P & R & F$_1$ & $\Delta$F$_1$ \\
    \midrule
    \texttt{zero\_shot} & --- & --- & --- & 0.158 & 0.7391 & 0.2604 & --- \\
    \texttt{taxonomy\_only} & \checkmark & --- & --- & 0.154 & 0.8478 & 0.2607 & 0.0003 \\
    \texttt{taxonomy\_rag} & \checkmark & \checkmark & --- & 0.1416 & 0.7174 & 0.2365 & -0.0242 \\
    \texttt{full} & \checkmark & \checkmark & \checkmark & 0.1692 & 0.7826 & 0.2783 & 0.0418 \\
    \bottomrule
  \end{tabular}
\end{table}

\subsection{Confirmation Outcomes by Class (RQ3)}
Of \NConfProcessed{} submitted findings, \NConfirmed{} were confirmed and all \NInconclusive{} remained inconclusive. The failure arose primarily before fuzzing: \NHarnessBuildFail{} harnesses did not compile, leaving only \NHarnessUsable{} (\PctHarnessUsable\%) usable attempts. Of these, \NNoCrash{} completed the \FuzzBudget{}-second budget without a sanitizer fault and \NHarnessBuilt{} compiled too late to execute. The failure taxonomy identifies the integration barrier. \NFailUndecl{} failures (\PctFailUndecl\%) involved undeclared types, \NFailSig{} involved signature mismatches, and \NFailNoMem{} referenced nonexistent members. These failures reflect missing access to generated message types, constructors, initialization order, and executor context. By contrast, only \NFailIncl{} failures involved missing includes and \NFailLink{} involved linking. The primary obstruction was therefore the ROS~2 type and initialization surface rather than the compiler or fuzzer.

This is a negative result about function-level confirmation methodology, not evidence that the reported weaknesses were absent. Single-process harness synthesis is poorly matched to middleware-dependent AV software when most attempts fail before execution.

\subsection{Producer Disagreement (RQ3)}
RQ3 concerns weaknesses that rule-based analyzers cannot structurally express. Of the \NClasses{} taxonomy classes, \NRuleExpr{} correspond to matchable code constructs; the remaining \NNonExpr{} depend on missing checks, semantic conventions, or deployment properties for which no syntax-tree pattern distinguishes safe from unsafe code, as summarized in Table~\ref{tab:taxonomy}. The producers separate along this boundary. The analyzers mapped \NToolMapped{} findings to \NToolClasses{} classes, \PctToolExpr\% of them in rule-expressible classes. Their remaining \NToolNonExpr{} findings fall in only \NToolNonExprClasses{} non-expressible classes: \NToolNonExprOne{} in unsafe degradation on component failure, from our best-effort Semgrep rules, and \NToolNonExprTwo{} in missing staleness validation, from cppcheck. The best-effort rules written for the other \NBestEffortSilent{} non-expressible classes produced nothing at all, which is a sharper result than declining to write them would have been. The model conditions mapped \NLLMMapped{} findings across all \NClasses{} classes, including \NLLMNonExpr{} (\PctLLMNonExpr\%) in non-expressible classes, and reached \NClassLLMOnly{} classes (\ClassLLMOnlyList{}) that no analyzer reached, all \NClassLLMOnlyNonExpr{} of them non-expressible by construction. Across all producers, \NClusters{} location clusters were reported, including \NClustersLLMOnly{} reported by only one model condition.

Because these findings were not dynamically confirmed and precision was not established, they remain candidates rather than verified weaknesses. The structural result is narrower: rule-based analyzers cannot directly represent the non-expressible classes, whereas model-based analysis can generate findings within them. Whether those findings are correct requires independent adjudication.


\section{Discussion}\label{sec:discussion}

The principal result is one of recall rather than precision. Model-based analysis localized weaknesses later corrected upstream at rates not explained by flag volume, including locations missed by all four static analyzers. However, the review value of the remaining findings is unknown because dynamic confirmation failed for reasons unrelated to finding validity. The results therefore do not support unattended deployment over an AV codebase.

The taxonomy produced the highest recall, by a margin too small to interpret from a single run, and raised explicit class labeling from near zero to above four-fifths. Its clearest contribution is therefore organizational: class-labeled findings can be prioritized, routed, and audited by vehicle-level consequence. Retrieval showed no benefit and should be reconsidered unless a corpus closer to ROS~2 application code is available. These findings support complementing, not replacing, static analysis. Rule-based tools are deterministic, do not fabricate, and are well suited to weaknesses carried by explicit code constructs, which accounted for \PctToolExpr\% of their findings. Model-based analysis is most relevant to the \NNonExpr{} non-expressible classes, whose signatures involve missing checks, semantic conventions, or deployment properties. Writing the AV-specific CodeQL queries illustrated the limitation directly: detecting flows from ROS~2 messages required sourcing taint at field reads, because it did not propagate from a \texttt{shared\_ptr} callback parameter through the middleware's dereference. Without explicit modeling of that idiom the queries returned nothing, despite more than twelve hundred subscription callbacks in the stack.

Dynamic confirmation requires message-level fuzzing against a live ROS~2 node graph, with harness synthesis supplying the relevant message definitions and constructors. Combined with blind manual triage, this would enable reliable estimates of precision and review cost.


\section{Conclusions}\label{sec:conclusions}

We introduced an AV vulnerability taxonomy of \NClasses{} operationally defined
classes and used it as machine-readable context in LLMSec-AV, evaluated on
\NUnitsEval{} Autoware code units against CodeQL, Semgrep, cppcheck, and the
Clang Static Analyzer, the first two extended with hand-written AV-specific
rules. On \NGroundTruth{} locations mined from upstream fixes, the model
conditions recovered up to \NullCodTaxObs{}. Any producer that reports densely
will match some locations by coincidence, so each recall figure was compared
against what that same producer would achieve if its own findings were scattered
at random over the code it reported. All \NLLMConditions{} model conditions beat
that expectation, by \MinLLMLift{} to \MaxLLMLift{}, and no analyzer did;
cppcheck matched one location while raising \NCppcheckFindings{} alerts. What the
models recovered therefore reflects where they looked, not how much they flagged.
The advantage lies primarily in model-based analysis, while the taxonomy provided
the highest recall by a small margin and increased explicit class labeling from
near zero to above four-fifths. Our study also identified a structural limitation
of rule-based tooling: \NNonExpr{} taxonomy classes are not expressible as static
rules, and the analyzers covered \NClassLLMOnly{} fewer classes than the model
conditions. Dynamic confirmation was constrained mainly by software build
integration, with \NHarnessBuildFail{} of \NConfProcessed{} attempts failing to
compile, largely because of undeclared types.

Although this study would be the first AV-specific vulnerability taxonomy used as machine-readable analysis context rather than a descriptive result, there are several limitations. These results remain bounded by incomplete ground truth and limited experimental scope. Precision is not established because findings outside the \NGroundTruth{} mined locations cannot be reliably classified, and the intended fuzzing-derived ground truth was empty. Our mined locations are heuristic, \NKnownMed{} of \NKnownIssues{} remain medium-confidence, and the \MatchTol{}-line tolerance may credit nearby but distinct findings. Baseline coverage with the traditional tools was also limited to \NBaseInUnits{} of \NBaseAll{} findings across \NBaseUnitsTouched{} of \NUnitsEval{} units. In addition, our study uses one public stack of AV, two quantized local models, one run per condition, and no variance estimate. The results therefore support analytic rather than statistical generalization, and small differences among ablation conditions (zero\_shot, taxonomy\_only, taxonomy\_rag, and full) should not be overinterpreted.

Future work should replace function-level harnessing of the source code with message-level fuzzing of live ROS~2 node graphs and provide harness synthesis with message definitions and constructor signatures. This confirmation substrate, together with blind manual triage, is necessary to establish precision of the findings. Replication of the analysis should also extend to another stack and middleware, such as Apollo, evaluate multiple seeds and stronger models, and focus on classes that cannot be represented as static-analysis rules. The most promising assurance architecture is therefore a domain-grounded machine-learning model layer that prioritizes AV-specific findings for review alongside conventional analyzers.


\section*{DATA AVAILABILITY}
The taxonomy, retrieval corpus manifest, prompt templates, analyzer configurations, and all evaluation artifacts are organized for release, with every reported quantity traceable to a committed artifact file.

\section*{ACKNOWLEDGMENTS}
We used the generative AI tools `ChatGPT' and `Claude' to help rephrase parts of our own writing to improve clarity and for editorial purposes.

\section*{AUTHOR CONTRIBUTIONS}
\textbf{Md. Wasiul Haque, Sagar Dasgupta:} conceptualization, methodology, coding, data collection, data analysis, and writing – original draft; \textbf{Mizanur Rahman:} conceptualization, methodology, writing – original draft, review and editing, and funding acquisition.

\section*{DECLARATION OF CONFLICTING INTERESTS}
The authors declared no potential conflicts of interest with respect to the research, authorship, and/or publication of this article.

\section*{FUNDING}
This research was supported by the National Center for Transportation Cybersecurity and Resiliency (TraCR) (a U.S. Department of Transportation National University Transportation Center) headquartered at Clemson University, Clemson, South Carolina, USA (Award \# 69A3552344812, 69A3552348317) and National Science Foundation (NSF) (Award \# 2340456). Any opinions, findings, conclusions, and recommendations expressed in this material are those of the author(s) and do not necessarily reflect the views of funding agencies, and the U.S. Government assumes no liability for the contents or use thereof.
\newpage
\bibliographystyle{trb}
\bibliography{main}

@inproceedings{koscher2010experimental,
  author    = {Koscher, Karl and Czeskis, Alexei and Roesner, Franziska and
               Patel, Shwetak and Kohno, Tadayoshi and Checkoway, Stephen and
               McCoy, Damon and Kantor, Brian and Anderson, Danny and
               Shacham, Hovav and Savage, Stefan},
  title     = {Experimental Security Analysis of a Modern Automobile},
  booktitle = {Proceedings of the IEEE Symposium on Security and Privacy},
  pages     = {447--462},
  year      = {2010}
}

@inproceedings{checkoway2011comprehensive,
  author    = {Checkoway, Stephen and McCoy, Damon and Kantor, Brian and
               Anderson, Danny and Shacham, Hovav and Savage, Stefan and
               Koscher, Karl and Czeskis, Alexei and Roesner, Franziska and
               Kohno, Tadayoshi},
  title     = {Comprehensive Experimental Analyses of Automotive Attack Surfaces},
  booktitle = {Proceedings of the 20th USENIX Security Symposium},
  year      = {2011}
}

@article{dieber2017ros,
  author  = {Dieber, Bernhard and Breiling, Benjamin and Taurer, Sebastian and
             Kacianka, Severin and Rass, Stefan and Schartner, Peter},
  title   = {Security for the Robot Operating System},
  journal = {Robotics and Autonomous Systems},
  volume  = {98},
  pages   = {192--203},
  year    = {2017}
}

@inproceedings{white2019sros,
  author    = {White, Ruffin and Christensen, Henrik I. and Caiazza, Gianluca
               and Cortesi, Agostino},
  title     = {Procedurally Provisioned Access Control for Robotic Systems},
  booktitle = {Proceedings of the IEEE/RSJ International Conference on
               Intelligent Robots and Systems (IROS)},
  pages     = {1--8},
  year      = {2018}
}

@inproceedings{mayoralvilches2022sros2,
  author    = {Mayoral-Vilches, V{\'i}ctor and White, Ruffin and
               Caiazza, Gianluca and Arguedas, Mikael},
  title     = {{SROS2}: Usable Cyber Security Tools for {ROS} 2},
  booktitle = {Proceedings of the IEEE/RSJ International Conference on
               Intelligent Robots and Systems (IROS)},
  pages     = {11253--11259},
  year      = {2022}
}

@article{vilches2018rvss,
  author  = {Mayoral Vilches, V{\'i}ctor and Gil-Uriarte, Endika and
             Zamalloa Ugarte, Irati and Olalde Mendia, Gorka and
             Izquierdo Pis{\'o}n, Rodrigo and Alzola Kirschgens, Laura and
             Bilbao Calvo, Asier and Hern{\'a}ndez Cordero, Alejandro and
             Apa, Lucas and Cerrudo, C{\'e}sar},
  title   = {Towards an Open Standard for Assessing the Severity of Robot
             Security Vulnerabilities, the Robot Vulnerability Scoring System
             ({RVSS})},
  journal = {arXiv preprint arXiv:1807.10357},
  year    = {2018}
}

@inproceedings{maggi2022dds,
  author    = {Maggi, Federico and Boasson, Erik and Cheng, Mars and
               Kuo, Patrick and Toyama, Chizuru and
               Mayoral Vilches, V{\'i}ctor and Vosseler, Rainer and
               Yen, Ta-Lun},
  title     = {The Data Distribution Service ({DDS}) Protocol Is Critical:
               Let's Use It Securely},
  booktitle = {Black Hat Europe},
  year      = {2022}
}

@inproceedings{cao2019adversarial,
  author    = {Cao, Yulong and Xiao, Chaowei and Cyr, Benjamin and Zhou, Yimeng
               and Park, Won and Rampazzi, Sara and Chen, Qi Alfred and
               Fu, Kevin and Mao, Z. Morley},
  title     = {Adversarial Sensor Attack on {LiDAR}-Based Perception in
               Autonomous Driving},
  booktitle = {Proceedings of the ACM SIGSAC Conference on Computer and
               Communications Security (CCS)},
  pages     = {2267--2281},
  year      = {2019}
}

@inproceedings{shin2017illusion,
  author    = {Shin, Hocheol and Kim, Dohyun and Kwon, Yujin and Kim, Yongdae},
  title     = {Illusion and Dazzle: Adversarial Optical Channel Exploits
               Against Lidars for Automotive Applications},
  booktitle = {Cryptographic Hardware and Embedded Systems (CHES)},
  pages     = {445--467},
  year      = {2017}
}

@inproceedings{garcia2020comprehensive,
  author    = {Garcia, Joshua and Feng, Yang and Shen, Junjie and
               Almanee, Sumaya and Xia, Yuan and Chen, Qi Alfred},
  title     = {A Comprehensive Study of Autonomous Vehicle Bugs},
  booktitle = {Proceedings of the ACM/IEEE International Conference on
               Software Engineering (ICSE)},
  pages     = {385--396},
  year      = {2020}
}

@inproceedings{lou2022study,
  author    = {Lou, Guannan and Deng, Yao and Zheng, Xi and Zhang, Mengshi and
               Zhang, Tianyi},
  title     = {Testing of Autonomous Driving Systems: Where Are We and Where
               Should We Go?},
  booktitle = {Proceedings of the 30th ACM Joint European Software Engineering
               Conference and Symposium on the Foundations of Software
               Engineering (ESEC/FSE)},
  pages     = {31--43},
  year      = {2022}
}

@techreport{iso21434,
  author      = {{International Organization for Standardization}},
  title       = {{ISO/SAE 21434:2021} Road Vehicles --- Cybersecurity
                 Engineering},
  institution = {ISO},
  year        = {2021}
}

@techreport{unece155,
  author      = {{United Nations Economic Commission for Europe}},
  title       = {{UN Regulation No.\ 155} --- Cyber Security and Cyber Security
                 Management System},
  institution = {UNECE},
  year        = {2021}
}

@misc{cwe,
  author       = {{MITRE Corporation}},
  title        = {Common Weakness Enumeration},
  howpublished = {\url{https://cwe.mitre.org/}},
  year         = {2024}
}

@misc{nvd,
  author       = {{National Institute of Standards and Technology}},
  title        = {National Vulnerability Database},
  howpublished = {\url{https://nvd.nist.gov/}},
  year         = {2024}
}

@misc{autoware,
  author       = {{Autoware Foundation}},
  title        = {Autoware: Open-Source Software for Autonomous Driving},
  howpublished = {\url{https://github.com/autowarefoundation/autoware}},
  year         = {2024}
}

@misc{ros2design,
  author       = {{Open Robotics}},
  title        = {{ROS} 2 Design: Executors and Callback Groups},
  howpublished = {\url{https://design.ros2.org/}},
  year         = {2024}
}

@article{macenski2022ros2,
  author  = {Macenski, Steven and Foote, Tully and Gerkey, Brian and
             Lalancette, Chris and Woodall, William},
  title   = {Robot Operating System 2: Design, Architecture, and Uses in the Wild},
  journal = {Science Robotics},
  volume  = {7},
  number  = {66},
  pages   = {eabm6074},
  year    = {2022}
}

@inproceedings{kato2018autoware,
  author    = {Kato, Shinpei and Tokunaga, Shota and Maruyama, Yuya and
               Maeda, Seiya and Hirabayashi, Manato and Kitsukawa, Yuki and
               Monrroy, Abraham and Ando, Tomohito and Fujii, Yusuke and
               Azumi, Takuya},
  title     = {Autoware on Board: Enabling Autonomous Vehicles with Embedded
               Systems},
  booktitle = {Proceedings of the ACM/IEEE International Conference on
               Cyber-Physical Systems (ICCPS)},
  pages     = {287--296},
  year      = {2018}
}

@techreport{omgdds,
  author      = {{Object Management Group}},
  title       = {Data Distribution Service ({DDS}), Version 1.4},
  institution = {OMG},
  year        = {2015}
}

@techreport{ieee754,
  author      = {{IEEE}},
  title       = {{IEEE} Standard for Floating-Point Arithmetic
                 ({IEEE} Std 754-2019)},
  institution = {IEEE},
  year        = {2019}
}

@techreport{saej3016,
  author      = {{SAE International}},
  title       = {Taxonomy and Definitions for Terms Related to Driving
                 Automation Systems for On-Road Motor Vehicles ({J3016})},
  institution = {SAE International},
  year        = {2021}
}

@inproceedings{fioraldi2020aflpp,
  author    = {Fioraldi, Andrea and Maier, Dominik and Ei{\ss}feldt, Heiko and
               Heuse, Marc},
  title     = {{AFL++}: Combining Incremental Steps of Fuzzing Research},
  booktitle = {Proceedings of the USENIX Workshop on Offensive Technologies
               (WOOT)},
  year      = {2020}
}

@inproceedings{serebryany2012asan,
  author    = {Serebryany, Konstantin and Bruening, Derek and
               Potapenko, Alexander and Vyukov, Dmitriy},
  title     = {{AddressSanitizer}: A Fast Address Sanity Checker},
  booktitle = {Proceedings of the USENIX Annual Technical Conference},
  year      = {2012}
}

@inproceedings{avgustinov2016ql,
  author    = {Avgustinov, Pavel and de Moor, Oege and Jones, Michael Peyton
               and Sch{\"a}fer, Max},
  title     = {{QL}: Object-Oriented Queries on Relational Data},
  booktitle = {Proceedings of the 30th European Conference on Object-Oriented
               Programming (ECOOP)},
  pages     = {2:1--2:25},
  year      = {2016}
}

@inproceedings{wei2022cot,
  author    = {Wei, Jason and Wang, Xuezhi and Schuurmans, Dale and
               Bosma, Maarten and Ichter, Brian and Xia, Fei and
               Chi, Ed H. and Le, Quoc V. and Zhou, Denny},
  title     = {Chain-of-Thought Prompting Elicits Reasoning in Large Language
               Models},
  booktitle = {Advances in Neural Information Processing Systems (NeurIPS)},
  year      = {2022}
}

@inproceedings{lewis2020rag,
  author    = {Lewis, Patrick and Perez, Ethan and Piktus, Aleksandra and
               Petroni, Fabio and Karpukhin, Vladimir and Goyal, Naman and
               K{\"u}ttler, Heinrich and Lewis, Mike and Yih, Wen-tau and
               Rockt{\"a}schel, Tim and Riedel, Sebastian and Kiela, Douwe},
  title     = {Retrieval-Augmented Generation for Knowledge-Intensive {NLP}
               Tasks},
  booktitle = {Advances in Neural Information Processing Systems (NeurIPS)},
  year      = {2020}
}

@inproceedings{zhou2019devign,
  author    = {Zhou, Yaqin and Liu, Shangqing and Siow, Jingkai and Du, Xiaoning
               and Liu, Yang},
  title     = {Devign: Effective Vulnerability Identification by Learning
               Comprehensive Program Semantics via Graph Neural Networks},
  booktitle = {Advances in Neural Information Processing Systems (NeurIPS)},
  year      = {2019}
}

@article{chakraborty2021deep,
  author  = {Chakraborty, Saikat and Krishna, Rahul and Ding, Yangruibo and
             Ray, Baishakhi},
  title   = {Deep Learning Based Vulnerability Detection: Are We There Yet?},
  journal = {IEEE Transactions on Software Engineering},
  volume  = {48},
  number  = {9},
  pages   = {3280--3296},
  year    = {2022}
}

@inproceedings{xia2024fuzz4all,
  author    = {Xia, Chunqiu Steven and Paltenghi, Matteo and
               Tian, Jia Le and Pradel, Michael and Zhang, Lingming},
  title     = {{Fuzz4All}: Universal Fuzzing with Large Language Models},
  booktitle = {Proceedings of the ACM/IEEE International Conference on Software
               Engineering (ICSE)},
  year      = {2024}
}

@misc{semgrep,
  author       = {{Semgrep, Inc.}},
  title        = {Semgrep: Lightweight Static Analysis for Many Languages},
  howpublished = {\url{https://semgrep.dev/}},
  year         = {2025}
}

@misc{cppcheck,
  author       = {Marjam{\"a}ki, Daniel},
  title        = {Cppcheck: A Tool for Static C/C\raisebox{0.15ex}{++} Code
                  Analysis},
  howpublished = {\url{https://cppcheck.sourceforge.io/}},
  year         = {2025}
}

@misc{clangsa,
  author       = {{LLVM Project}},
  title        = {Clang Static Analyzer},
  howpublished = {\url{https://clang-analyzer.llvm.org/}},
  year         = {2025}
}

@misc{codestral,
  author       = {{Mistral AI}},
  title        = {Codestral: A Generative Model for Code Generation},
  howpublished = {\url{https://mistral.ai/news/codestral/}},
  year         = {2024}
}

@misc{gptoss,
  author       = {{OpenAI}},
  title        = {{gpt-oss}: Open-Weight Reasoning Models},
  howpublished = {\url{https://openai.com/index/introducing-gpt-oss/}},
  year         = {2025}
}

@misc{apollo,
  author       = {{Baidu}},
  title        = {Apollo: An Open Autonomous Driving Platform},
  howpublished = {\url{https://github.com/ApolloAuto/apollo}},
  year         = {2025}
}

@article{chen2021codex,
  author  = {Chen, Mark and Tworek, Jerry and Jun, Heewoo and Yuan, Qiming and
             Pinto, Henrique Ponde de Oliveira and Kaplan, Jared and
             Edwards, Harri and Burda, Yuri and Joseph, Nicholas and
             Brockman, Greg and others},
  title   = {Evaluating Large Language Models Trained on Code},
  journal = {arXiv preprint arXiv:2107.03374},
  year    = {2021}
}

@inproceedings{fan2023llmse,
  author    = {Fan, Angela and Gokkaya, Beliz and Harman, Mark and
               Lyubarskiy, Mitya and Sengupta, Shubho and Yoo, Shin and
               Zhang, Jie M.},
  title     = {Large Language Models for Software Engineering: Survey and Open
               Problems},
  booktitle = {IEEE/ACM International Conference on Software Engineering:
               Future of Software Engineering (ICSE-FoSE)},
  pages     = {31--53},
  year      = {2023}
}

@techreport{iso26262,
  author      = {{International Organization for Standardization}},
  title       = {{ISO 26262:2018} Road Vehicles --- Functional Safety},
  institution = {ISO},
  year        = {2018}
}

@incollection{haque2027security,
  title={Security Vulnerabilities in Software Supply Chain for Autonomous Vehicles},
  author={Haque, Md Wasiul and Erfan, Md and Dasgupta, Sagar and Rahman, Md Rayhanur and Rahman, Mizanur},
  booktitle={ADVANCES IN TRANSPORTATION CYBERSECURITY AND RESILIENCY},
  pages={97--146},
  year={2027},
  publisher={World Scientific}
}

@article{haque2026llm,
  title={LLM-Assisted Dynamic Threat Analysis for Attacker-Reachable Software Weaknesses in Autonomous Vehicles},
  author={Haque, Md Wasiul and Dasgupta, Sagar and Rahman, Mizanur and Rahman, Md Rayhanur},
  journal={arXiv preprint arXiv:2608.13450},
  year={2026}
}
\end{document}